# Adaptive flexibility of cells through nonequilibrium entropy production

Yuika Ueda[1] and Shinji Deguchi[1,*]

[1] Division of Bioengineering, Graduate School of Engineering Science, Osaka University

[*] Corresponding author

Address: 1-3 Machikaneyama, Toyonaka, Osaka 560-8531, Japan

E-mail: deguchi.shinji.es@osaka-u.ac.jp

Phone: +81 6 6850 6215

ORCID: 0000-0002-0556-4599

# Abstract

Cellular adaptation to environmental changes relies on the dynamic remodeling of subcellular structures. Among these, sarcomere structures are fundamental to the organization and function of the cytoskeletal architecture. In muscle-type cells, sarcomeres exhibit ordered structures of consistent lengths, optimized for stable force generation. By contrast, nonmuscle-type cells display a higher degree of structural variability, with sarcomeres of varying lengths that contribute not only to force generation but also to adaptive remodeling upon environmental cues. While these differences in sarcomere structures have traditionally been attributed to the unique properties of specific proteins expressed in each cell type, the functional implications of such structural variability remain unclear. Here, we present a nonequilibrium physics framework to elucidate the role of sarcomere variability in cytoskeletal adaptation. Specifically, we demonstrate that the effective binding strength of sarcomere components can be evaluated by analyzing structural randomness using Shannon entropy. The increased entropy associated with the inherent randomness of sarcomere structures in nonmuscle-type cells lowers the energy barrier for cytoskeletal remodeling, enabling flexible adaptation to environmental demands. Meanwhile, the ordered sarcomere arrangements in muscle-type cells correspond to higher binding energies and more stable cytoskeletal configurations. Although structural disorder is often regarded as unfavorable in terms of stability, our study suggests that it plays a key role in enabling adaptive responses in cellular systems.

# 1. Introduction

Living cells adapt their internal structures in response to changes in both their internal and external environments (1,2). This adaptability is fundamental to individual cellular processes such as differentiation, proliferation, and apoptosis, as well as to higher-order processes such as tissue development and wound healing (3,4). The cytoskeleton is central to the physical structure and force generation of cells, maintaining its integrity through organized protein complexes. Among these, the sarcomere, a periodic structure primarily composed of actin and myosin II filaments, functions as the fundamental unit of the cellular contractile structure.

While sarcomeres are well defined in muscle cells (5–9), similar structures are also found in nonmuscle cells (10–18). Compared to the consistent sarcomere length and alignment in muscle cells, nonmuscle sarcomeres exhibit greater structural variability; and even within muscle cells, younger ones display more randomness than mature ones (19,20). This variability is more pronounced in motile cells than in nonmotile cells, which possess different potentials for movement and proliferation (11,12). Muscle sarcomeres are optimized for contractile force generation within muscle tissues (5), while nonmuscle sarcomeres are involved in diverse dynamic cellular processes such as division, migration, and structural remodeling upon environmental changes (21–24).

Traditionally, these differences in sarcomere structures have been attributed to the unique properties of specific proteins expressed in each cell type, often within a reductionist framework. On the other hand, to achieve a unified understanding of how this architectural randomness is inherently linked to distinct cellular functions, a comparative approach based on physical principles is essential. In this regard, prior studies from a physical perspective have aimed to elucidate the mechanisms for the self-organization of sarcomere periodicity (20–22,25,26), with some attempts in exploring how ATP-driven actomyosin contractions generate cellular tension and in turn maintain sarcomere structures (5,27,28). However, the potential link between sarcomere variability and cellular functions including adaptability remains unclear.

Here, we develop a nonequilibrium physics framework to describe the relationship between sarcomere structural variability and cellular adaptability. Specifically, we derive constraints on the probability distribution of sarcomere structures and the effective binding energy among sarcomeres to evaluate how structural variability can enhance adaptability. Given that ATP-dependent processes represent nonequilibrium phenomena, we incorporate biophysical properties of sarcomeres into a Fokker–Planck equation to derive the probability distribution of sarcomere configurations. We then quantify entropy production to assess the energetic cost of maintaining these structures. In addition, we analyze the sarcomere length distribution from experimental results on different cell types and clarify its functional relationship in our framework. Our results reveal a fundamental physical principle, demonstrating that the force-bearing subcellular structure, often considered only at the scale of individual contractile units, enhance adaptive flexibility at larger scales of cells. Interestingly, the inherent randomness in sarcomere structure, which impairs function in engineered systems, instead promotes the unique adaptability of living organisms.

# 2. Model

## 2.1 Probability distribution of sarcomere structures

The sarcomere is the intracellular contractile unit with a certain periodic structure, along which actin, myosin, and α-actinin appear in a cell type-dependent probability distribution (Fig. 1). Nonmuscle-type cells tend to express sarcomeres that are more spatially irregular compared to those in muscle-type cells. We model the sarcomere in a one-dimensional framework, defining the probability of its length state $x$ as $P(x)$. Given the biophysical property of the sarcomere response (see Supporting note 1), which is convex downward with respect to the stable length $x_0$, the potential energy $U(x)$ at sarcomere length $x$ is simply given by

$$U(x) = \frac{1}{2}k(x - x_0)^2 \tag{1}$$

where $k$ represents the restoring contribution of the sarcomere. Using the Fokker-Planck equation, the time evolution of $P(x,t)$ is described by

$$\frac{\partial P(x,t)}{\partial t} = -\frac{\partial}{\partial x}\{-k(x-x_0)P(x,t)\} + \frac{\partial^2}{\partial x^2}\{DP(x,t)\} \tag{2}$$

where $D$ is the diffusion coefficient. By solving Eq. (2) under the steady-state condition, we obtain the stationary probability distribution as

$$P(x) = \frac{\exp\left(-\frac{k(x-x_0)^2}{2D}\right)}{\sum \exp\left(-\frac{k(x-x_0)^2}{2D}\right)} \tag{3}$$

where $P(x)$ satisfies the normalization condition $\sum P(x) = 1$.

## 2.2 Entropy production

We consider the cytoskeletal structure with sarcomeres as the system and the surrounding environment as the heat bath. We quantify the nonequilibrium entropy production during the growth process of the cytoskeletal structure while nearby sarcomeres bind together or unbind. In actual cells, the cytoskeletal growth occurs at the level of individual constituent proteins (29), namely mainly actin, myosin, and α-actinin, rather than at the whole sarcomere level, but we simplify this process by modeling the combined action of these proteins as equivalent to sarcomere level binding. We assume that the environment is large enough to remain close to thermal equilibrium. The total entropy production, considered for both the system and the environment, is defined as

$$\sigma = \Delta S_{sys} + \Delta S_{bath} \geq 0 \tag{4}$$

where $\Delta S_{sys}$ and $\Delta S_{bath}$ are the change in entropy of the system and of the surrounding environment, respectively (30).

In nonequilibrium systems, $\Delta S_{sys}$ can be expressed using Shannon entropy as follows (26):

$$\Delta S_{sys} = -\sum P(x) \ln P(x). \tag{5}$$

For $\Delta S_{bath}$, we assume that the environment can freely exchange constituent sarcomere elements and energy with the preexisting cytoskeletal structure and is therefore described by a grand canonical distribution under this theoretical assumption. The change in entropy of the environment is given by

$$\Delta S_{bath} = -\beta(\Delta E - \Delta\mu) \tag{6}$$

where $\Delta E$ is the energy change in the cytoskeletal structure due to sarcomere binding, $\Delta\mu$ is the change in chemical potential as sarcomeres transition from the particle bath to within the cytoskeletal structure, and $\beta = 1/k_b T$ representing the inverse thermal energy. Using Eqs. (5) and (6), the total entropy production is expressed by

$$\sigma = -\sum P(x)\ln P(x) - \beta(\Delta E - \Delta\mu) \geq 0, \tag{7}$$

showing that the entropy production during the forming process of the cytoskeletal structure is always positive.

## 2.3 Maximum binding energy

We define $\Delta E$ in Eq. (6) as the energy change due to sarcomere binding to the cytoskeletal structure, corresponding to the binding energy, or $\Delta E = \Delta E^b$. From Eq. (7), the binding energy is defined as

$$\Delta E^b \leq \Delta E^b_{max} = \Delta\mu - \frac{1}{\beta}\sum P(x)\ln P(x). \tag{8}$$

Based on the second law of thermodynamics, the nonequilibrium forming of the cytoskeletal structure driven by sarcomere binding consistently generates positive entropy. This constraint imposes a limit on the binding energy, which represents the strength of the interaction between sarcomeres, with a more negative binding energy indicating a stronger binding. We quantitatively analyzed the maximum binding energy $\Delta E_{max}^b$ with the probability distribution of sarcomere structures described by Eq. (3). Each parameter used is determined based on experimental data, with $D = 10$ μm²/s and $T = 310.15$ K (31–35), and $\Delta\mu$ was estimated to be on the order of 20 kJ/mol from experimental measurements of the energy associated with ATP hydrolysis (36,37).

## 2.4 Quantification of sarcomere length

The actual sarcomere length was quantified using previously published experimental images from multiple sources, including mouse skeletal muscles (38), primary mice myotubes (39), neonatal rat cardiomyocytes (40), and myofibrils isolated from rabbit glycerinated psoas muscle fibers in a rigor state (41), which predominantly express striated muscle-type actin and myosin molecules; and human fibroblasts (42), PtK2 long-nosed potoroo epithelial kidney cells (42), gerbil fibroma cells (42), and A7r5 rat aortic smooth muscle cells (43), which express nonstriated muscle-type actin and myosin molecules. Among them, A7r5 cells also express α-SMA, a smooth muscle-specific actin isoform commonly used as a marker for smooth muscle cells and myofibroblasts. Images of endogenous α-actinin immunostaining (39,40,42,43), transmission electron microscopy images (39), or phase-contrast microscopy images (41) were analyzed by using ImageJ/Fiji software (NIH). Lines were drawn along the longitudinal axis of multiple sarcomeres, and the intensity profile was examined to measure the distance between peaks, which correspond to sarcomere length. The lengths of $n$ = 40–48 sarcomeres in striated muscle types and $n$ = 53–56 sarcomeres in nonstriated muscle types were analyzed to obtain the respective distributions. To illustrate representative sarcomere distributions, we also hereby performed immunostaining for α-actinin and actin filaments on A7r5 cells following method (43) and on human foreskin fibroblasts following method (29) (Fig. 1).

# 3. Results

## 3.1 Analysis of sarcomere length variability

The probability distribution for sarcomere lengths was analyzed using our model (Eq. 3) (Fig. 2). These distributions follow a Gaussian distribution with variance $\sqrt{D/k}$, influenced by both noise and restoring contribution of sarcomeres. As $k$ increases, representing greater contractile force maintained within the cytoskeletal structure, the variance decreases, resulting in sarcomeres being more stably distributed around a specific length. Conversely, as $k$ decreases, the variance increases, leading to a broader length distribution and a more random sarcomere structure. This behavior is consistent with experimentally observed properties in muscle and nonmuscle cells. Specifically, in mouse skeletal muscles (hereafter referred to as Skeletal muscles), primary mice myotubes (Myotubes), neonatal rat cardiomyocytes (Cardiomyocytes), and myofibrils isolated from rabbit glycerinated psoas muscle fibers in a rigor state (Skeletal myofibrils), which all predominantly express striated muscle-type actin and myosin molecules and are characterized by high contractile force, sarcomeres are consistently observed to maintain a specific length (Fig. 3a); to clearly demonstrate this observation, Shannon entropy was calculated for each distribution (Fig. 3b). In contrast, in human fibroblasts (Fibroblasts), PtK2 long-nosed potoroo epithelial kidney cells (PtK2 cells), and gerbil fibroma cells (Fibroma cells), which express nonstriated muscle-type actin and myosin molecules and are characterized by less contractile force, sarcomeres exhibit greater length variability. Thus, nonmuscle sarcomeres show a relatively random structure compared to those in muscle cells, while still maintaining a characteristic stable length (15,44). A7r5 rat aortic smooth muscle cells (A7r5 cells), which express both α-SMA and muscle-type α-actin, exhibit intermediate randomness among the analyzed cell types.

## 3.2 Analysis of maximum binding energy

Our analysis shows that the maximum binding energy decreases as the restoring contribution $k$ increases, indicating that the impact of the binding energy associated with the extent of sarcomere interactions increases under these conditions (Fig. 4a). This occurs because an increased restoring contribution leads to a more ordered sarcomere structure, thus reducing the influence of entropy (Fig. 2, blue) and in turn allowing the binding energy to exert greater dominance in driving cytoskeletal elongation in Eq. (8). On the other hand, when the restoring contribution is small, the resulting more disordered sarcomere configuration (Fig. 2, red) reduces the dominance of the binding energy, allowing sarcomere assembly to occur more easily and enabling more flexible cytoskeletal remodeling. These findings suggest that cytoskeletal structures with random sarcomere arrangements, as observed in nonmuscle cells, are better adaptable for dynamic elongation. In contrast, cytoskeletal structures with ordered sarcomeres, such as those in muscle cells, require greater changes in binding energy for elongation, suggesting that these systems are less suited for dynamic remodeling. Interestingly, the value of $x_0$ does not affect the limits of the binding energy because entropy is determined not by the absolute sarcomere stable length but by the probability of the structure being randomly distributed (Fig. 4a).

The relationship with the chemical potential reveals a positive correlation with the maximum binding energy as explicitly described in Eq. (8) (Fig. 4b). A small difference in chemical potential (Fig. 4b, red regions) corresponds to a reduced impact of the binding energy, allowing for easier remodeling of the structures. While our discussion has focused on the extent of the sarcomere randomness, it should be noted that chemical potential depends on component concentration, which varies with intracellular environmental conditions. Therefore, even with the same randomness in the sarcomere system, the behavior is actually influenced by environmental factors including concentration.

The effective binding energy for different types of sarcomere structures was estimated using Eq. (8) and actual measurements of the distribution (Fig. 3), showing its proportionality to the change in system entropy at a fixed chemical potential, -20 kJ/mol (Fig. 5). The nonmuscle sarcomeres tend to have lower levels of binding energy compared to muscle ones, indicating that the former can remodel their structures more easily than the latter.

## 4. Discussion

In this work, we developed a nonequilibrium physical model to elucidate how the biophysical properties of sarcomeres influence cytoskeletal stability (Fig. 4). Previous studies have largely focused on self-organizing mechanisms of sarcomere patterning (21,22,25,26), but it remained unclear how sarcomere characteristics are associated with cellular adaptation to environmental changes. We modeled the probability distribution of sarcomere length by integrating essential physical factors such as restoring mechanisms and fluctuations into an energy model. Using the Fokker-Planck equation, we derived the probability distribution of sarcomere lengths (Eq. (3)), revealing that higher restoring forces induce less random structural variations around a specific length. This finding is consistent with experimental observations, in which muscle cells exhibit ordered sarcomere structures, whereas nonmuscle cells display more dispersed configurations (5,10–17,45,46). Interestingly, cytoskeletal stiffness increases with cellular aging, accompanied by reduced sarcomere variabilities (47), potentially reflecting a decline in cellular adaptability during maturation as cells become increasingly specialized for contraction.

We determined the entropy production during cytoskeletal remodeling due to sarcomere binding. Sarcomeres generate contractile forces via ATP-driven actomyosin activity, creating a nonequilibrium system with continuous energy flux. In our model, we considered the cytoskeletal structure composed of sarcomeres as the system and its surrounding environment as the reservoir. Given that sarcomere binding allows for cytoskeletal elongation, we modeled the environment as a grand canonical ensemble and quantified the resulting entropy production. By determining Shannon entropy (48) from the probability distribution of sarcomere lengths, we derived a limit on binding energy (Eq. (8)), constrained by the requirement of nonnegative entropy production in nonequilibrium systems (30). Our analysis then provided a mechanism, by which the maximum binding energy is influenced by structural randomness. In muscle-type cells with high restoring force and ordered sarcomere lengths, the requirement for entropy production necessitates substantial binding energy for cytoskeletal elongation. In contrast, in nonmuscle-type cells with lower restoring force and more random sarcomere lengths, the entropy resulting from sarcomere disorder allows for more flexible cytoskeletal remodeling with minimal binding energy. Thus, these cytoskeletal structures achieve flexible responsiveness to environmental cues facilitated by the randomness-driven entropy. Interestingly, young muscle cells display greater randomness in sarcomeres than mature cells (19), potentially suggesting that early-stage cells may be more adaptable and then gradually form specialized structures optimized for stable contractile force generation as they mature. Note that binding energy is also modulated by the chemical potential, which depends on intracellular component concentrations. For example, in high ATP environments, stable cytoskeletal structures with strong binding energies can form even in the presence of random sarcomere structures. Thus, cytoskeletal structures can remodel in response to conditions within the system as well as those in the environment. Our theoretical descriptions capture this cellular behavior, aligning with experimentally observed relationships across the different hierarchies, namely between sarcomeres and cells (5,10–17,45).

While we have primarily discussed the qualitative features, we also estimated the effective binding energy for different types of sarcomere structures (Fig. 5). We found that the estimated tendency is consistent with experimental observations. Specifically, nonmuscle α-actinin isoforms have been reported to exhibit higher dissociation constants $K_d$ for actin binding, ranging from 2.96 to 3.96 μM (49), compared to muscle α-actinin isoforms, which display lower $K_d$ values of 0.4 μM (50) or 0.59 μM (51). This difference suggests that muscle α-actinin binds actin more stably, stabilizing the sarcomere more effectively than nonmuscle α-actinin. Regarding myosin, namely another major actin cross-linker, the $K_d$ value of muscle myosin isoforms (28.2 nM) is higher than that of nonmuscle myosin ones (4.6 nM) (52), suggesting a higher affinity of nonmuscle myosin for actin. Functionally, muscle myosin molecules undergo rapid cross-bridging cycles to facilitate muscle contraction, whereas nonmuscle myosin molecules remain attached to actin to sustain cellular tension along stress fibers. These distinct roles suggest that α-actinin, rather than myosin, predominantly drives the remodeling activity of sarcomere structures.

In conclusion, we developed a physical framework linking different hierarchical processes, namely subcellular structural randomness and cellular adaptability. We suggest that structural disorder, often perceived as disadvantageous, actually enhances cellular adaptability by increasing entropy. This randomness is not limited to sarcomeres but also appears inherently in other cellular components, indicating that extending our model could reveal additional mechanisms working across other biological hierarchies. Thus, our framework provides a basis for exploring the complex relationships between cellular physical structures and functions.

## Acknowledgments

We thank Tsubasa S. Matsui and Shiyou Liu for their technical assistance related to Fig. 1. Y.U. is supported by the Japan Society for the Promotion of Science (JSPS). This study was supported in part by JSPS KAKENHI grants (23H04929 and 24KJ1649).

# Figures

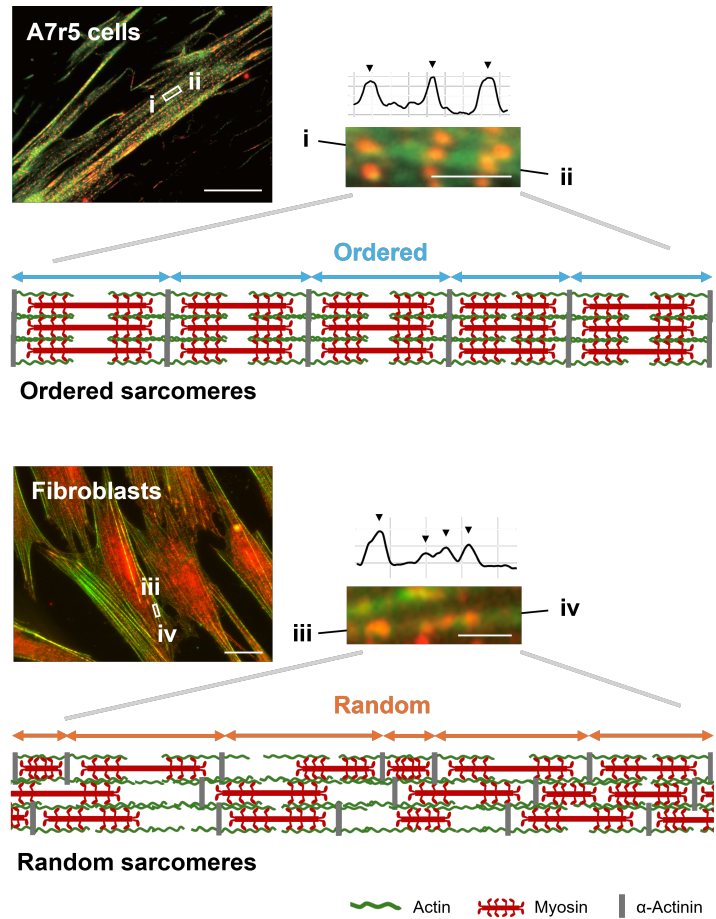

Fig. 1  Schematic of sarcomere structure. The upper diagram (blue) shows ordered sarcomeres with specific lengths, while the lower diagram (orange) shows random sarcomeres with varying lengths. White rectangles are magnified to evaluate the intensity of α-actinin (red) along the indicated stress fibers (green). Arrow heads indicate peaks in the intensity curves. Scale, 20 μm (left) and 2 μm (magnified).

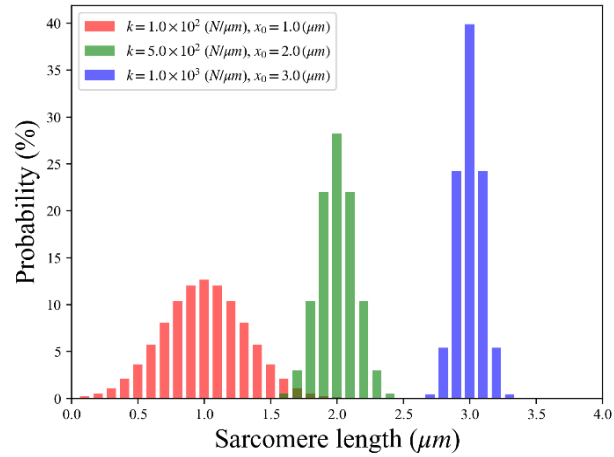

Fig. 2 Length distribution of sarcomeres at $k = 10^2$ [N/μm] (red), $10^3$ [N/μm] (green), and $10^4$ [N/μm] (blue). With a smaller $k$, the distribution of sarcomere lengths exhibits greater variability, indicating increased structural diversity.

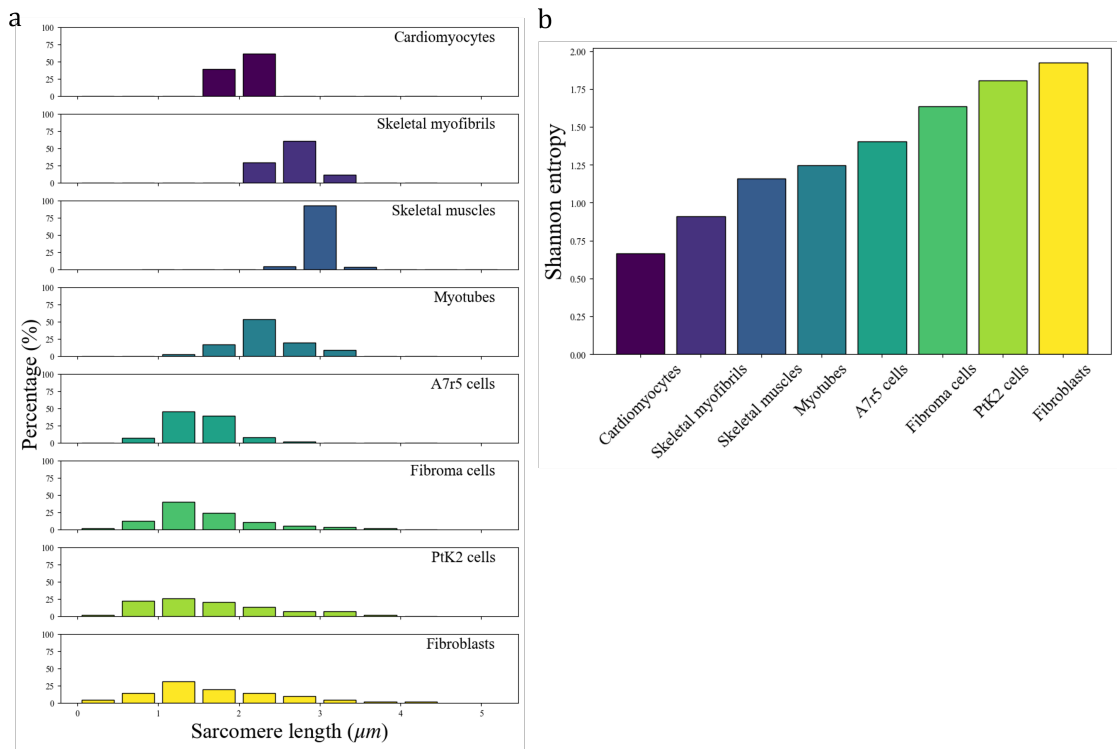

Fig. 3 Quantification of sarcomere length variability. (a) The actual sarcomere length was quantified using previously published experimental images from multiple sources. (b) Shannon entropy was calculated for each distribution.

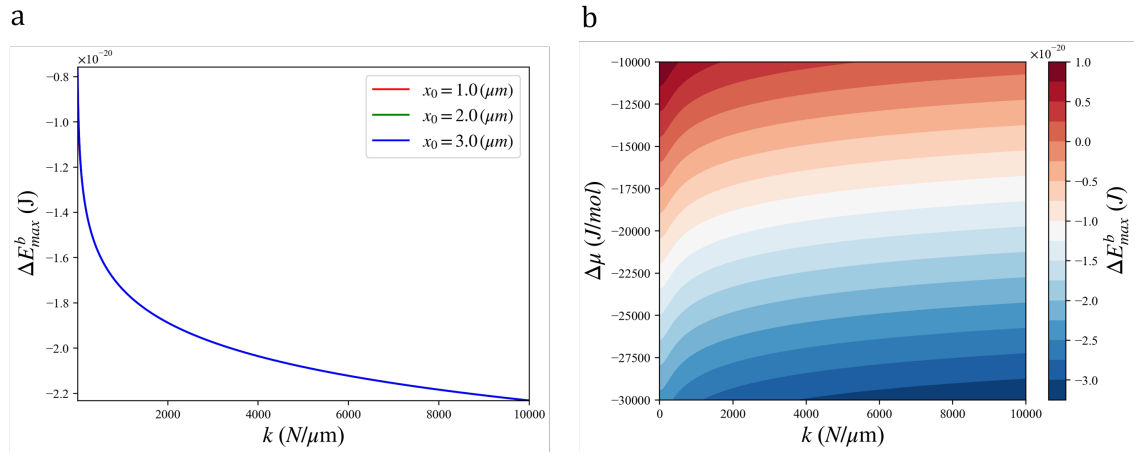

Fig. 4 Quantification of maximum binding energy. (a) Relation between restoring contribution and maximum binding energy. A higher restoring contribution corresponds to a lower maximum binding energy, indicating stronger binding between constituent sarcomere elements. (b) Dependence of maximum binding energy on restoring contribution and chemical potential. A lower chemical potential results in a lower maximum binding energy.

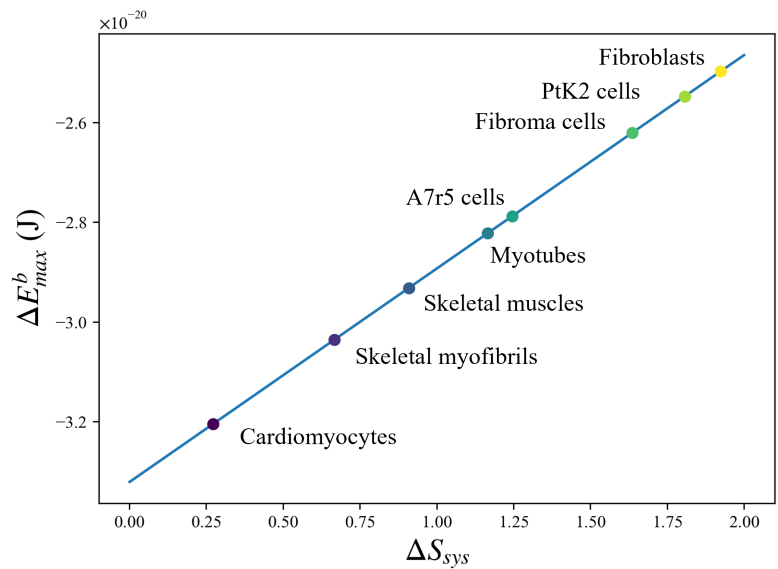

Fig. 5 Estimation of binding energy for different sarcomere structures. The binding energy for different types of sarcomere structures was estimated using Eq. (8) and the distributions shown in Fig. 3 at a fixed chemical potential of -20 kJ/mol.

**Ueda and Deguchi, "Adaptive flexibility of cytoskeletal structures through nonequilibrium entropy production"**

**Supporting note 1 - Thermodynamic explanation of the stability of sarcomeres**

The main text conceptualizes the formation process of sarcomere-based cytoskeletal structures within a nonequilibrium framework. In contrast, Supporting note 1 describes the potential energy of an individual sarcomere within a closed-system framework, representing equilibrium conditions where key sarcomere components, i.e., actin, myosin, and α-actinin, are already assembled and isolated from environmental interactions. This distinction between equilibrium and nonequilibrium frameworks reflects the processes being analyzed. Specifically, it highlights whether we focus on the dynamic interactions of sarcomeres with the external environment, as in a nonequilibrium framework, or its behavior of a preexisting sarcomere in isolation, as in an equilibrium framework. Here, we describe the basis for the restoring contribution introduced in the main text, showing how deviations from a specific sarcomere length appear to generate elastic restoring forces and allowing for the maintenance of a stable sarcomere length.

We assume that changes in the length of sarcomeres are caused by the displacement of the actin filaments (AFs; Fig. S1, green) along the axial direction. Myosin filaments (MFs; red) and AFs interact within the sarcomere across four distinct regions (top-left, bottom-left, top-right, and bottom-right of the MFs), and we consider these interactions to exhibit symmetrical changes across these regions. We define the following four chemical potentials: the chemical potential of AF–MF regions where MFs and AFs are bound, denoted as $\mu_{acmy}^{co}$; the chemical potential of AF regions where MFs and AFs are unbound because of the absence of MFs, denoted as $\mu_{ac}^{sb}$; the chemical potential of MF regions where MFs and AFs are unbound because of the overlap between them, denoted as $\mu_{my}^{sb}$; and the chemical potential of AF regions where AFs overlap with each other, denoted as $\mu_{ac}^{l}$. It is known that once unbound AFs and MFs are mixed, the reaction spontaneously progresses toward binding, and thus the bound state of AFs and MFs is energetically favorable. Consequently, the chemical potentials satisfy the relations $\mu_{acmy}^{co} < \mu_{ac}^{sb}$ and $\mu_{acmy}^{co} < \mu_{my}^{sb}$. The free energy $G$ is expressed as $G = H - TS = \sum \mu_j N_j$, where $H$ is enthalpy, $T$ is thermodynamic temperature, $S$ is entropy, $\mu_j$ is chemical potential of species $j$, and $N_j$ is the number of constituent particles of $j$,. The entropy of non-overlapping AFs should be greater than that of overlapping AFs. Furthermore, since actin filaments (AFs) do not directly bundle with each other, the enthalpy change between them does not affect the free energy change. Thus, the free energy difference for AFs at constant temperature is determined as $\Delta G = \sum \mu_i \Delta N_i$, which implies that the chemical potential satisfies $\mu_{ac}^{sb} < \mu_{ac}^{l}$ because the free energy of non-overlapping AFs is lower.

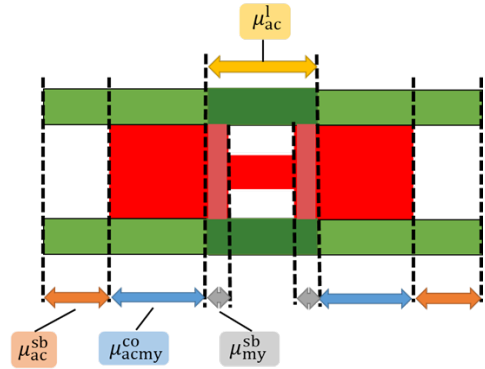

Fig. S1 Sarcomere model defining four different chemical potentials. MFs (red) exists between AFs (green).

We examine the change in free energy when AFs shift by $\Delta x$ relative to MFs, resulting in a total change of $2\Delta x$ across the entire sarcomere. Focusing on the binding states of MFs and AFs, as well as the overlap between MFs, we classify the sarcomere into four distinct characteristic states (Fig. S2). Using changes in free energy, we show the direction of changes in sarcomere length. First, consider the elongated state of the sarcomere caused by external tension (Fig. S2a). In this state, the unbound regions of MFs and AFs increase as the sarcomere lengthens. Denoting a reactant with chemical potential $\mu_i$ as $[\mu_i]$, the reaction can be expressed as

$$4\Delta x[\mu_{ac}^{sb}] + 2\Delta x[\mu_{my}^{sb}] \leftrightarrows 2\Delta x[\mu_{acmy}^{co}] \quad (S1)$$

The change in free energy, considering the relative magnitudes of the chemical potentials described earlier, is

$$\Delta G = 2\Delta x\{\mu_{acmy}^{co} - (2\mu_{ac}^{sb} + \mu_{my}^{sb})\} < 0 \tag{S2}$$

Thus, the reaction proceeds in a direction where AFs move inward to increase their binding with MFs, allowing the sarcomere length to shorten. Next, we consider the state where MFs are sufficiently bound to AFs, and no overlap exists between AFs (Fig. S2b). The reaction in this case is expressed as

$$4\Delta x[\mu_{ac}^{sb}] \leftrightarrows 4\Delta x[\mu_{ac}^{sb}] \tag{S3}$$

The change in free energy is

$$\Delta G = 4\Delta x(\mu_{ac}^{sb} - \mu_{ac}^{sb}) = 0 \tag{S4}$$

As the free energy change in this case is zero, the system is in equilibrium, and thus the sarcomere can be considered to have reached a stable length. Next, when the sarcomere is in a state where MFs are sufficiently bound to AFs, but AFs overlap due to contraction (Fig. S2c), the reaction is expressed as

$$4\Delta x[\mu_{ac}^{sb}] \leftrightarrows 4\Delta x[\mu_{ac}^{l}] \tag{S5}$$

The free energy change in this case is

$$\Delta G = 4\Delta x(\mu_{ac}^{l} - \mu_{ac}^{sb}) > 0 \tag{S6}$$

Thus, the reaction proceeds outward, elongating the sarcomere. Finally, the sarcomere is in a state where AFs overlap, and the overlap interferes with MF-AF binding (Fig. S2d), the

reaction is expressed as

$$4\Delta x[\mu_{ac}^{sb}] + 2\Delta x[\mu_{acmy}^{co}] \leftrightarrows 4\Delta x[\mu_{ac}^{l}] + 2\Delta x[\mu_{my}^{sb}] \tag{S7}$$

The corresponding free energy change is

$$\Delta G = 2\Delta x\{2(\mu_{ac}^{l} - \mu_{ac}^{sb}) - (\mu_{my}^{sb} - \mu_{acmy}^{co})\} \tag{S8}$$

Thus, the reaction proceeds outward, elongating the sarcomere.

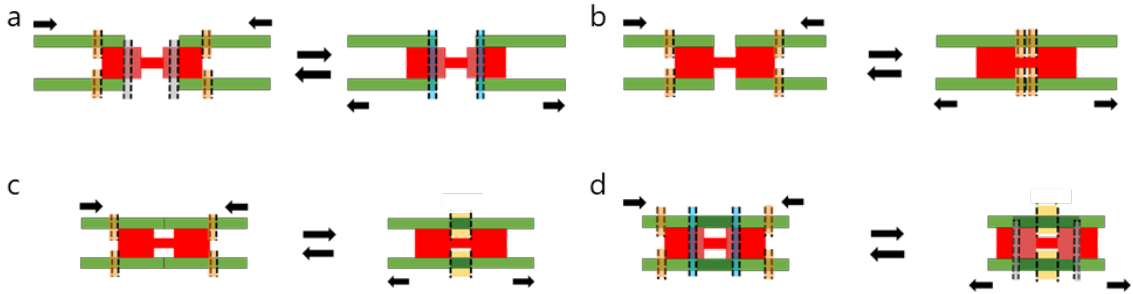

Fig. S2 Different states of sarcomere. (a–d) Elongated (a), normal (b), slightly shortened (c), and significantly shortened states (d).

As demonstrated above, the direction of sarcomere length changes is determined by the free energy change (Fig. S3). When MFs and AFs are sufficiently bound, and no overlap exists between AFs, the sarcomere reaches its lowest free energy state, corresponding to the stable sarcomere length $x_0$ in the main text. When the sarcomere is shorter or longer than its stable length, it spontaneously elongates or shortens toward stability, respectively. In a state where MFs and AFs are fully bound and there is no overlap between AFs, the myosin heads are expected to perform efficient power strokes, thereby generating maximum contractile force. Thus, we can interpret that sarcomere responds to length change to have a potential energy that is convex downward with respect to the stable length as described by

Eq. (1) in the main text.

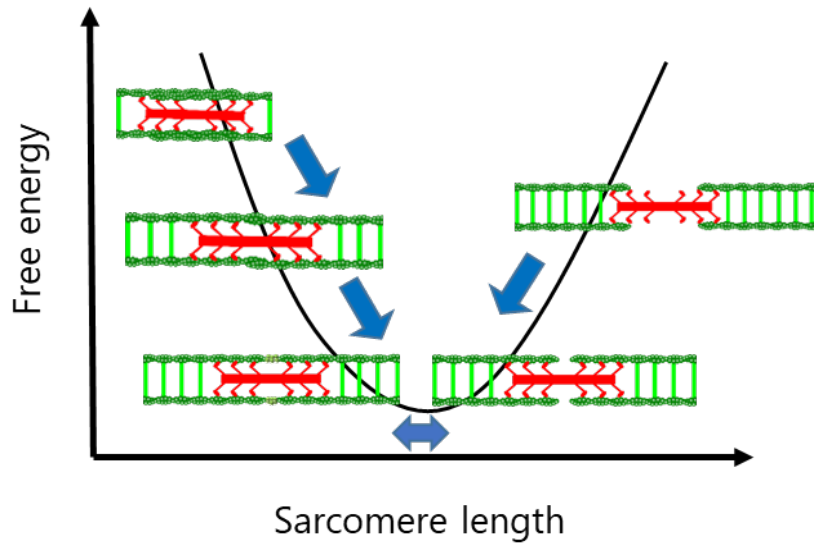

Fig. S3 Schematic of the free energy change as a function of sarcomere length. Blue arrows show the direction of spontaneous change in sarcomere length.